\documentstyle[aps,multicol,epsf]{revtex}

\begin{document}

\title{
Anomalous percolating properties of growing 
networks
}

\author{
S.N. Dorogovtsev$^{1, 2,}$\footnote{Electronic address: sdorogov@fc.up.pt}, 
J.F.F. Mendes$^{1,}$\footnote{Electronic address: jfmendes@fc.up.pt}, 
and 
A.N. Samukhin$^{2,}$\footnote{Electronic address: alnis@samaln.ioffe.rssi.ru} 
}

\address{
$^{1}$ Departamento de F\'\i sica and Centro de F\'\i sica do Porto, Faculdade 
de Ci\^encias, 
Universidade do Porto\\
Rua do Campo Alegre 687, 4169-007 Porto, Portugal\\
$^{2}$ A.F. Ioffe Physico-Technical Institute, 194021 St. Petersburg, Russia 
}

\maketitle

\begin{abstract} 
We describe the anomalous phase transition of the emergence of the giant connected component in scale-free networks growing under mechanism of preferential linking. 
We obtain exact results for the size of the giant connected component 
and the distribution of vertices among connected components. 
We show that all the derivatives of the giant connected component size $S$ over 
the rate $b$ of the emergence of new edges 
are zero at the percolation threshold $b_c$, and 
$S \propto \exp\{-d(\gamma)(b-b_c)^{-1/2}\}$, where the coefficient $d$ is a function of the degree distribution exponent $\gamma$. 
In the entire phase without the giant component, these networks are in a ``critical state'':  
the probability ${\cal P}(k)$ that a 
vertex belongs to a connected component of a size $k$ is of a power-law form. 
At the phase transition point, ${\cal P}(k) \sim 1/(k\ln k)^2$. 
In the phase with the giant component, 
${\cal P}(k)$ has an exponential cutoff at 
$k_c \propto 1/S$.
In the simplest particular case,  
we present exact results for growing exponential networks.   
\end{abstract}


\begin{multicols}{2}

\narrowtext


\section{Introduction}
\label{intr}

From a physical point of view, networks may be equilibrium and non-equilibrium. For example, to the class of equilibrium networks belong classical random graphs with randomly distributed connections introduced by Erd\"os and R\'enyi \cite{er59,er60} and their generalizations \cite{mr95,mr98,acl00,bck01}. Percolating properties of equilibrium networks are well-studied \cite{mr95,mr98,nw99i,mn00,nsw00,cebh00,cnsw00,cebh01,dm01b}. 
The behavior of these networks near the threshold point, that is, near the point of the emergence of the giant connected component, is similar to percolation on an infinite-dimensional 
lattice.  

From the other hand, the most important real networks (the Internet and the World Wide Web, for instance) have the growing total numbers of the vertices and, thus, are non-equilibrium \cite{ajb99,ba99,ajb00,s01,bkm00,t01}. 
The growth of networks, which is often a self-organization process, 
produces a number of intriguing effects \cite{ba99,bb00a,bb00b}. 

Very recently, it was found that the percolation transition in growing networks is of a quite different nature than in equilibrium ones \cite{chkns01}. 
(Note that the notions of percolation and a giant connected component are meaningful only in the large network limit.) 
In Ref. \cite{chkns01}, for the growing network with an exponential degree distribution (degree is the number of connections of a vertex), it was found numerically that this transition is of an infinite order. 
All the derivatives of the size of the giant connected component (the percolating cluster) are zero at the percolation threshold. 

Here we propose a theory of the anomalous percolation transition in 
growing networks including the most interesting and important growing scale-free networks. 
Also, as a particular case, the exponential growing networks are considered.
Thus, we present the complete exact description of the percolation transition both for the scale-free and exponential growing networks. These cases have been turned 
to be similar to each other, and the phase transition is of an infinite order. 
Furthermore, we show that {\em in the entire phase without the giant connected component such growing networks are in a ``critical state''}. 
In this state, the probability ${\cal P}(k)$ that a randomly chosen vertex belongs to a connected component of the size $k$ is of a power-law form.

\section{Model and degree distribution}
\label{model} 

For constructing the scale-free network, we apply the mechanism of preferential attachments of new edges \cite{ba99,s55,sbook57,f41,s43}. Here we use one of the simplest models producing power-law degree distributions \cite{dm00c}: 

(i) At each increment of time, a new vertex is added to the network, so that 
the total number of vertices in the network is $t$.  

(ii) Simultaneously, $b$ new undirected links are distributed between vertices according to the following rule. The probability that a new edge connects a pair $(\mu,\nu)$ of vertices is proportional to $(q_\mu+a)(q_\nu+a)$, where $q_\mu$ and $q_\nu$ are the degrees of these vertices, and $a$ is some positive constant which plays the role of additional attractiveness of vertices for new edges \cite{dms00}. $b$ is also an arbitrary positive constant. Multiple edges are forbidden (in principle, they are non-essential for large networks). 

The degree distribution $P(q)$ of the network (degree is the total number of connections of a vertex) can be easily obtained using standard considerations (for example, see Refs. \cite{dms00,krl00,kr00,krr00,kk00,kk01}). It is of the form 

\begin{equation}
P(q) = \left(1 + \frac{a}{2b} \right) 
\frac{\Gamma(a+1+\frac{a}{2b})}{\Gamma(a)} 
\frac{\Gamma(q+a)}{\Gamma(q+a+2+\frac{a}{2b})}
\, ,  
\label{1}
\end{equation}  
where $\Gamma(\ )$ is the gamma-function. 
For large $q$, $P(q) \propto q^{-\gamma}$, where $\gamma=2+a/(2b)$. 
It is convenient to introduce a new notation 
$\zeta \equiv \gamma-2 = a/(2b)$.  
In the limit $\zeta \to \infty$, the $\gamma$ exponent approaches $\infty$, and one can check that the degree distribution turns to be exponential.

\section{Evolution of connected components}
\label{evolution}

According to the above rules, a new vertex may have no connections, so disjoint vertices and components are certainly present in the network. We focus on the distribution of the sizes of connected components (clusters of mutually connected vertices) and 
on the size of the giant connected component. 
How do {\em finite} connected components grow with time? 
Here we present an elementary consideration of the evolution of connected components. For a rigorous derivation of our main equations, see Appendix \ref{ao}. 
For the large network, it is almost impossible that both the ends of a new edge are being attached to the same finite connected component. 
Let us use the following essential circumstance. 
{\em In the large network,the finite size connected components are almost surely trees.} 
Obviously, this is not the case for the giant connected component.) 
This fact is the key issue of percolation theory for networks \cite{mr95,mr98,nsw00}. 
One can check that the number of edges in a finite connected component (tree) with $k$ vertices is equal to $k-1$. Therefore, the total degree of this component equals $2(k-1)$, and the probability that a new edge is attached to the component is proportional to $2k-2+ka$. This value should be normalized. 
Taking into account that the total degree of the network is equal to $2b t$, we find that this probability is $[(2+a)k-2]/[(2b+a)t]$. 

The resulting equation for the number $N_k(t)$ of connected components of size $k$ at time $t$ looks like 

\begin{eqnarray}
&& N_k(t+1) = \delta(k-1) + 
\left[ 1 - 2b\,\frac{(2+a)k-2}{(2b+a)t} \right] N_k(t) + 
\nonumber
\\[5pt]
&& b \sum_{j=1}^{k-1} 
\frac{(2+a)j-2}{(2b+a)t}\, \frac{(2+a)(k-j)-2}{(2b+a)t}\, N_j(t) N_{k-j}(t) 
\, ,  
\label{2}
\end{eqnarray}    
where $\delta(k-1)$ is a convenient notation for the Kronecker symbol. 
The first term on the right-hand side of Eq. (\ref{2}) accounts for the emergence of new vertices. The second term describes the decrease of the number of connected components of the size $k$ due to attaching of new edges to the vertices of these components. The last term on the right-hand side of Eq. (\ref{2}) accounts for the fusion of connected components into larger ones of the size $k$. Here again we have used the largeness of the network to present the terms of the sum in the factorized form (see Ref. \cite{dm00c}). 
The limit $a \to \infty$ corresponds to the absence of any preference. 
In this case, new edges connect the pairs of randomly chosen vertices, and Eq. (\ref{2}) takes the form of the master equation derived in Ref. \cite{chkns01}.  
We emphasize that Eq. (\ref{2}) is nonlinear unlike master equations for degree distribution \cite{dms00,krl00}.

\section{Main equations}
\label{main}

When $t \to \infty$, the ratio $N_k(t)/t$ approaches the stationary value $n_k$. The equation for it is of the form 

\begin{eqnarray}
&& \left[ 1 + 2b\frac{2+a}{2b+a}\left(k - \frac{2}{2+a} \right) \right] n_k = \delta(k-1) +
\nonumber
\\[5pt]
&& b\! \left(\frac{2+a}{2b+a} \right)^{\!\!2}\, 
\sum_{j=1}^{k-1} \left(j - \frac{2}{2+a} \right)\! \left(k-j - \frac{2}{2+a} \right)\! n_j n_{k-j} 
,  
\label{3}
\end{eqnarray}  
Our main matter of interest is the probability ${\cal P}(k)$ that a 
randomly chosen 
vertex belongs to a connected component with $k$ vertices, that is 
${\cal P}(k) = kn_k$. 
Let us introduce its $Z$-transform:  
$g(z) \equiv \sum_{k=1}^{\infty} kn_k z^k$. 
Similarly, 
$n(z)  \equiv \sum_{k=1}^{\infty} n_k z^k$. 
To approach our aim, that is the description of the connected component statistics and the giant connected component, we must find $g(z)$. 
One sees that $g(z) = z n^\prime (z)$ and $n(z) = \int_0^z dz g(z)/z$. 
Then, from Eq. (\ref{3}) (see also Appendix \ref{ao}), we obtain the equation 

\begin{eqnarray}
&& g(z) - z + 2b\frac{2+a}{2b+a} 
\left(zg^\prime(z) - \frac{2}{2+a}g(z) \right) \times  
\nonumber
\\[5pt]
&& \left\{1 - \frac{2+a}{2b+a} \left[g(z) - \frac{2}{2+a}n(z)\right] \right\} = 0
\, .  
\label{4}
\end{eqnarray} 
If the giant connected component is absent, that is, $g(1)=1$, all the connected components are almost surely trees, so 
$\sum_k [(2+a)k-2]N_k = (2b+a)t$, 
i.e., $(2+a)g(1) - 2n(1) = 2b+a$, 
and we obtain the necessary condition:  
$n(1)=1-b$ when $g(1)=1$. One can easily check that Eq. (\ref{4}) satisfies this condition.  

Using the convenient combination

\begin{equation}
h(z) \equiv 1 - 
\frac{2+a}{2b+a}g(z) + \frac{2}{2b+a}\int_0^z \frac{dz}{z}g(z)
\, ,   
\label{5}
\end{equation}  
we rewrite Eq. (\ref{4}) in the form 

\begin{equation}
g(z) - z - b z \frac{d}{dz}h^2(z) = 0
\, .   
\label{6}
\end{equation} 
Therefore, 

\begin{equation}
n(z) = \int_0^z \frac{dz}{z}g(z) = b[h^2(z)-1] + z
\, .   
\label{7}
\end{equation}   
Substituting Eqs. (\ref{6}) and (\ref{7}) into Eq. (\ref{5}), we finally obtain the equation 

\begin{equation}
(2+a)z h(z)h^\prime(z) - h^2(z) + (1+\zeta)h(z) - \zeta(1-z) = 0
\, ,   
\label{8}
\end{equation}     
where $\zeta \equiv \gamma-2 = a/(2b)$. One sees that $g(0)=n(0)=0$, so 
Eq. (\ref{8}) must be supplied with the boundary condition $h(0)=1$. 
From Eqs. (\ref{6}) and (\ref{7}), a simple relation between $g(z)$ and $h(z)$ follows: 

\begin{equation}
g(z) = \frac{a}{2+a}
\left[ 
1 + \frac{2}{a}\,z + \frac{1}{\zeta}h^2(z) - \frac{1+\zeta}{\zeta}h(z) 
\right]
\, .   
\label{9}
\end{equation} 

From Eq. (\ref{8}), using Eq. (\ref{9}), we can obtain the $Z$-transform $g(z)$ of the basic distribution ${\cal P}(k)$ for finite connected components. 
When the giant connected component is absent, $g(1)=1$ and $n(1)=1-b$ as shown above. Then, from Eq. (\ref{5}), it follows that $h(1)=0$.   
As usually \cite{nsw00}, if the giant connected component exists, its size $S$ can be related to $g(1)$: $S = 1-g(1)$. 
After the introduction of new variable and function, 
$y \equiv z^{-1/(2+a)}$ and 
$\varphi(y)/y \equiv h(z)/(1+\zeta)$, 
Eq. (\ref{9}) takes the canonical form of the Abel equation of the second kind 

\begin{equation}
\varphi^\prime(y)\varphi(y) - \varphi(y) = \frac{\zeta}{1+\zeta} (y^{-1-a} - y)
\, .   
\label{10}
\end{equation}    
The boundary condition for Eq. (\ref{10}) corresponding to the condition $h(0)=1$ is 
$\varphi(y) \to y/(1+\zeta)$ as $y \to \infty$. 

Setting $y=1$ in Eq. (\ref{10}), we see that two situations are possible. 
If $\varphi(1)=0$, the giant connected component is absent. 
When $\varphi(1)\neq 0$, the giant connected component is present, and 
$\varphi^\prime(1)=1$.
Then, one can obtain from Eqs. (\ref{6}) or (\ref{9}) the expression for the size of the giant connected component

\begin{equation}
S = 1-g(1) = 
\frac{a}{2+a}\frac{(1+\zeta)^2}{\zeta}
\, \varphi(1)[1 - \varphi(1)]
\, .   
\label{12}
\end{equation}  
Thus, our problem is reduced to the analysis of the solutions to Eq. (\ref{10}).

\section{Phase diagram}
\label{phase} 

\begin{figure}
\epsfxsize=78mm
\epsffile{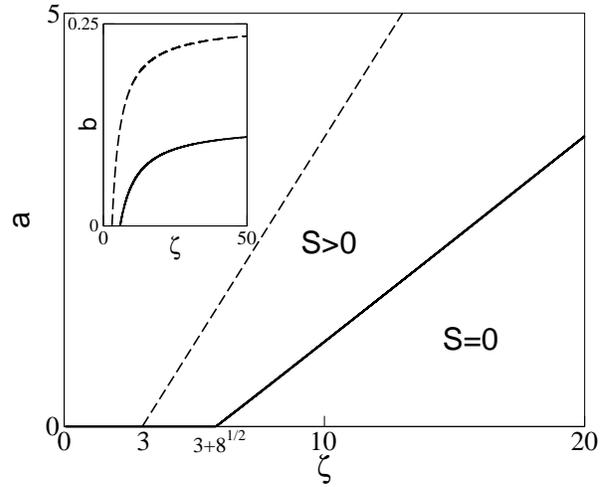}
\caption{
Phase diagram of the growing scale-free network. Here, $a$ is additional attractiveness of vertices for new edges, and $\zeta=\gamma-2=a/(2b)$, where $b$ is the value of the input flow of new edges. The solid line [see Eq. (\protect\ref{13})] indicates the phase transition of the emergence of the giant connected component. When $\zeta<3+2\sqrt2$, the giant connected component is absent ($S=0$) only on the $a=0$ line. 
The dashed line shows the percolation threshold line for the equilibrium random graph with the same degree distribution (\protect\ref{1}) as the growing scale-free network under consideration. 
Inset: the same phase diagram but plotted on axes $(\zeta,b)$.   
}
\label{f1}
\end{figure}

The simplest problem we must solve is to indicate the 
region of the parameters 
$b$ and $a$ or, equivalently, $\zeta$ and $a$ where the giant connected component is present. 
The direct analysis of Eq. (\ref{10}) (see Appendix \ref{aa}) yields the following picture (see Fig. \ref{f1}). When $\zeta > \zeta^\ast \equiv 3+2\sqrt{2}$, 
the giant connected component is absent below 
the phase transition line 

\begin{equation}
a(\zeta) = \frac{1}{4}(\zeta+\zeta^{-1}) - \frac{3}{2}
\, .   
\label{13}
\end{equation}  
For $\zeta<\zeta^\ast 
$, 
the trivial phase transition line is $a=0$.
The absence of the giant connected component at $a=0$ is obvious.
Indeed, when $\zeta \equiv a/(2b)$ is fixed, from $a \to 0$ follows $b \to 0$. In turn, zero input flow of edges produces a set of disjoint vertices.

For comparison, in Fig. \ref{f1}, we show the percolation threshold line for the equilibrium random graph with the same degree distribution (\ref{1}) as our growing scale-free network [see the dashed line $a(\zeta) = (\zeta-3)/2$]. This follows from the Molloy-Reed criterion for the existence of the giant connected component in equilibrium random graphs,  
$\sum_{q=0}^\infty (q^2-2q)P(q) > 0$ \cite{mr95,mr98}.

\section{Critical behavior}
\label{crit} 

\subsection{Size of the giant component}
\label{size} 

\begin{figure}
\epsfxsize=80mm
\epsffile{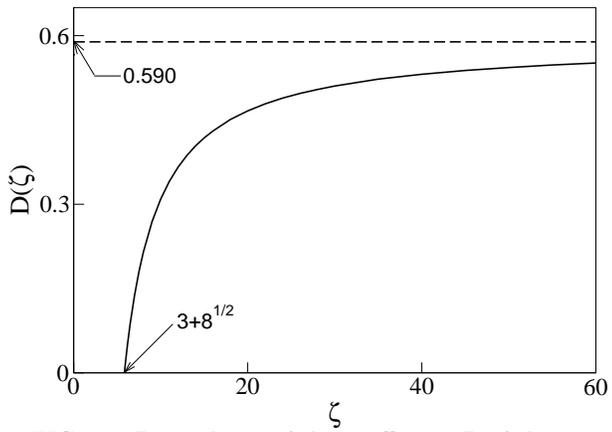}
\caption{
Dependence of the coefficient $D$ of the exponent in Eq. (\protect\ref{14}) on $\zeta$. Here, $\zeta=\gamma-2=a/(2b)$.
}
\label{f2}
\end{figure}

In equilibrium networks, the size of the giant connected component linearly approaches zero at the phase transition line. In growing networks we have a quite different situation. As shown in Appendix \ref{aa}, the size of the giant connected component near the phase transition line takes the form 

\begin{equation}
S(\zeta,a) = D(\zeta) \exp
\left\{-\frac{\pi}{2}\left[
\frac{\zeta(2+a)}{(1+\zeta)^2} - \frac{1}{4} \right]^{-1/2} \right\}
\, .   
\label{14}
\end{equation} 
The dependence of the factor $D$ on $\zeta$ is plotted in Fig. \ref{f2}. 
One sees that, in the case of the network with an exponential degree distribution,  
i.e., when $\zeta \to \infty$, the factor $D$ 
tends to a constant value $0.590\ldots$. On the other side, $D$ linearly approaches zero at $\zeta=
\zeta^\ast
$. 

In the following, to simplify our expressions, we shall use the notation

\begin{equation}
\lambda \equiv 
\left|  \frac{\zeta(2+a)}{(1+\zeta)^2} - \frac{1}{4} \right|^{1/2}
\, .   
\label{14a}
\end{equation} 
For the exponential network, i.e., when $\zeta \to \infty$, 
we have $\lambda = \sqrt{2|b-1/8|}$.

All the derivatives of the giant connected component size over the deviation from the critical line are zero. In particular, let us consider the deviation of $b$ from the critical line $b_c=b_c(\zeta)$ [the form of $b_c(\zeta)$ follows from Eq. (\ref{13})]. In this case, 
$S(\zeta,b)$ is of  the form 

\begin{equation}
S(\zeta,b) = D(\zeta) 
\exp
\left\{-\frac{\pi}{2\sqrt{2}} \frac{1+\zeta^{-1}}{\sqrt{b-b_c}} 
 \right\}
\, .   
\label{15}
\end{equation} 
In the limit of the exponential network, $\zeta \to \infty$, 
where $b_c=1/8$, we obtain 

\begin{equation}
S(\zeta \to \infty,b) = 0.590\ldots 
\exp
\left\{-\frac{\pi}{2\sqrt{2}} \frac{1}{\sqrt{b-b_c}} 
 \right\}
\, .   
\label{16} 
\end{equation} 
In Appendix \ref{ac}, we present a simple direct derivation of Eq. (\ref{16}). 
The factor in the index of the exponent is $\pi/(2\sqrt{2})=1.111\ldots$ that is in agreement with the result of numerics in Ref. \cite{chkns01}. 

\begin{figure}
\epsfxsize=75mm
\epsffile{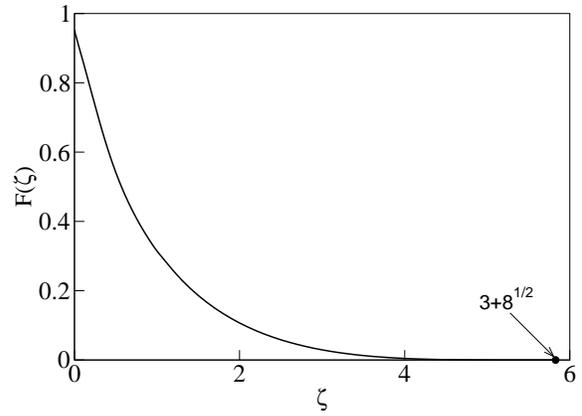}
\caption{
Factor $F$ in Eq. (\protect\ref{17}) vs $\zeta$. $F(0) = 0.95\ldots$.
}
\label{f3}
\end{figure}

For small $a$ and $0 < \zeta < \zeta^\ast
$, that is, near the other transition line ($a=0$, $0 < \zeta < \zeta^\ast$) on the phase diagram, the size of the giant connected component behaves in the following way 

\begin{equation}
S(\zeta,a) = F(\zeta) a
\, .   
\label{17} 
\end{equation} 
The factor $F$ versus $\zeta$ is plotted in Fig. \ref{f3}. All the derivatives of $F$ over $\zeta$ are zero at the point $\zeta = \zeta^\ast
$. 
Near this point, when $a$ is small and $\zeta \equiv \zeta^\ast
+\epsilon$, 
Eq. (\ref{15}) takes the form 

\begin{equation}
S(\epsilon,a) = 0.631\ldots a
\exp
\left\{-\sqrt{2}\pi[a - 4(3\sqrt{2}-4)\epsilon]^{-1/2}
 \right\}
\, .   
\label{18} 
\end{equation} 
Equation (\ref{18}) is valid when $a - 4(3\sqrt{2}-4)\epsilon>0$. 
Thus we see that the phase transition of the emergence of the giant connected component in growing networks is in sharp contrast to that in equilibrium nets. We shall discuss the nature of this anomalous phase transition in Sec. \ref{interp}. 

We close this subsection by the exact result for the relative total number $\ell$ of loops in the network (the ratio of the total number of loops in the network to its size $t$) obtained in the Appendix \ref{ao}. Note that all the loops are in the giant component since finite components are tree-like. The equation for $\ell$ looks like 

\begin{equation} 
\ell = \frac{1}{2}\,\frac{1}{1+\zeta}[(2+a)S + 2\ell]^2
\, .   
\label{18a} 
\end{equation} 
Here we do not present its trivial solution. Near the phase transition, where the giant component is small, $\ell \cong [(2+a)S]^2 /[2(1+\zeta)]$.

\subsection{Distribution of vertices among connected components}
\label{distr} 

The distribution of vertices among connected components is one of the 
key issues in percolation theory. In Appendix \ref{aa}, we calculate the probability ${\cal P}(k)$ that a randomly chosen vertex belongs to a connected component of the size $k$. 
 
At the point of the emergence of the giant connected component, we obtain 
\begin{equation}
{\cal P}(k) \cong  
\frac{2a}{2+a} \, \frac{1}{k^2 \ln^2 k}
\, .   
\label{19} 
\end{equation} 
Here, $a=a(\zeta)$ is given by the relation (\ref{13}) for the critical line. 
The factor $2a/(2+a)$ in Eq. (\ref{19}) approaches zero at the point 
$\zeta=\zeta^\ast
$. Equation (\ref{19}) must be compared with the corresponding result for percolation on the equilibrium networks and infinite-dimensional 
lattices, where ${\cal P}(k) \propto k^{-5/2}$ at the percolation threshold \cite{mr95,mr98,nsw00,sa91}. 

Furthermore, we find that in the entire phase without the giant connected component, ${\cal P}(k)$ is of a power-law form. 
In this phase, far from the phase transition, i.e., when the parameter $\lambda$ is not small [see the definition (\ref{14a})], 

\begin{equation}
{\cal P}(k) \sim k^{-4\lambda/(1-2\lambda)}
\, .   
\label{20} 
\end{equation}  
In the same phase, near the phase transition line (\ref{13}), ${\cal P}(k)$ has a power-law tail 

\begin{equation}
{\cal P}(k) = \frac{32a}{2+a}\lambda^2 [(2+a)k]^{-2-4\lambda}
\,    
\label{21} 
\end{equation}  
for $\ln k \gg 1/\lambda$, and coincides with the threshold distribution 
(\ref{19}) in the region $1 \ll \ln k \ll 1/\lambda$.

This power-law form is in striking contrast to the exponentially decreasing ${\cal P}(k)$ both above and below the percolation threshold for standard percolation   
\cite{sa91} including percolation on equilibrium networks. Equation (\ref{19}) indicates that the growing network is in a ``critical state'' in the entire phase without the giant connected component. Note that this is valid both for the scale-free and exponential growing networks. 
 
In the phase with the giant connected component, ${\cal P}(k)$ has an exponential tail. 
Near the phase transition, for the large values of connected component sizes 
$k \gg [a/(2+a)] S^{-1}$, we obtain the following behavior:

\begin{equation}
{\cal P}(k) = [2\pi (2ae)^3(2+a)S^{-3} k]^{-1/2} 
\exp [- \frac{2+a}{2ae}Sk]
\, .   
\label{22} 
\end{equation} 
Here, $a=a(\zeta)$ is taken for the transition line (\ref{13}), 
and $S=S(a,\zeta)$ is the size of the giant component given by 
Eq. (\ref{14}); $e = 2.73\ldots$. 
For smaller $k$, namely $1 \ll \ln k \ll 1/\lambda$, ${\cal P}(k)$ takes the form of the threshold distribution (\ref{19}). 
For $\ln k \sim 1/\lambda$, a crossover regime is present.

\section{Interpretation}
\label{interp} 

All the results of Secs. \ref{phase} and \ref{crit} were obtained by the explicit formal analysis of the master equations for growing networks. Let us discuss the physical nature of the behavior observed above. 

To begin with, let us recall that in networks growing under mechanism of preferential attachment of edges to vertices degree distributions of vertices are of a power-law (fractal) form. This means that such networks self-organize into scale-free structures with the power-law degree distributions while growing. In fact, they are in a critical state in a wide range of the values of the network parameters. 
The growth under mechanism of the preferential attachment produces power-law distributions. 
In principle, this phenomenon can be called self-organized criticality. 

In the present paper, we are interested not in the statistics of vertices but in the statistics of connected components, that is, in the distributions of the number of connections of distinct connected components and the distributions of the number of vertices in them. 
New links are being attached to large connected components with higher probability, so that large connected components have a better chance to merge and grow. This produces the preferential growth of large connected components even in networks where new edges are attached to randomly chosen vertices, that is, in networks without preferential attachment of edges to vertices. 
Such mechanism of the effective preferential attachment of new vertices to large connected components naturally produces power-law distributions of the sizes of connected components and power-law probabilities ${\cal P}(k)$. This ``self-organized critical state'' is realized in the growing networks only if the 
giant component is absent. 

As soon as the giant connected component emerges, the situation changes radically. A new channel of the evolution of the connected components is coming into play, and, with high probability, large connected components do not grow up to even larger ones but join to the giant component. Therefore, there are few large connected components if the giant component is present, and then ${\cal P}(k)$ is exponential. 

Thus, in the growing networks, two phases are in contact at the point of the emergence of the giant connected component --- the critical phase without the giant component and the normal phase with the giant component. This contact provides the above observed effects. There exists another example of a contact of a ``critical phase'' (or of a line of critical points) with a normal phase, namely, the Berezinskii-Kosterlitz-Thouless phase transition \cite{b70,kt73}. 
Interestingly, functional dependences in both these cases have similar functional forms. This indicates that equations describing such phase transitions have similar analytical properties. 

Connections in non-equilibrium networks are very in-homogeneously distributed between vertices. We mean that many edges are captured by old vertices, and few edges are attached to more young vertices --- ``the rich gets richer''. 
Nevertheless, this statistically in-homogeneous distribution of connections in growing networks is not the direct origin of the observed behavior. 
Both this in-homogeneity and the critical 
(or, one can say, power-law, or fractal, or scale-free) 
distributions of connected components in the absence of 
the giant component have the same first cause --- 
the specific process of the network growth.

\section{Conclusions}
\label{concl}

In summary, we have presented the theory of percolation in evolving systems --- growing networks. We have demonstrated that the interplay of a self-organization process and percolation produces a number of intriguing effects in such objects. We have obtained exact results for the size of the giant component and the distribution of vertices over connected components. An explicit description for the anomalous phase transition of the emergence of the giant component in these growing networks has been proposed. We hope that our results are of a general nature and can be applied to various growing systems.

\acknowledgments 
S.N.D. thanks PRAXIS XXI (Portugal) for a research grant PRAXIS XXI/BCC/16418/98. S.N.D. and J.F.F.M. 
were partially supported by the project POCTI/1999/FIS/33141. 
A.N.S. acknowledges the NATO program OUTREACH for
support. 
We also thank P.L. Krapivsky for useful discussions. 
\\


\appendix
\section{Rigorous description of the evolution of connected components}
\label{ao} 

In Secs. \ref{evolution} and \ref{main}, we have derived our main equation (\ref) using a simple but rather heuristic approach. Here we present a strict derivation of the equations describing the growth of the network. 

Let ${\cal N}(t)$ be the number of vertices in the growing network at time $t$. 
We assume that, with the probability $p(t)dt$, a new vertex is added to the network during a small time interval $dt$, i.e., ${\cal N}(t+dt)={\cal N}(t)+1$. The degree of a new vertex is supposed to be zero. 

The total degree of the network is 

\begin{equation}
{\cal Q}(t) = 2{\cal L}(t) = \sum_{i=1}^{{\cal N}(t)}q_i(t) 
\, ,   
\label{ao1} 
\end{equation}
where ${\cal L}(t)$ is the total number of edges in the network. 
We assume that with probability $b(t)$, a new edge emerges between vertices. According to the rule of preferential linking that we use in the present paper, the probability that this edge connects vertices $i$ and $j$ is equal to 

\begin{equation}
\frac{[q_i(t)+a][q_j(t)+a]}{[{\cal Q}(t)+a{\cal N}(t)]^2}
\,    
\label{ao2} 
\end{equation}
for each pair of vertices. 

If $b(t)=const$, one can set $p(t)=const=1$, so, in this case, we have two parameters, namely $b$ and $a$ --- additional attractiveness, that determine the growth and the structure of the network. 
Let us introduce a new object which we call the connectivity matrix of the network: 

\begin{equation}
S_{ij}(t) = \left\{ 
\begin{array}{rl}
1, &  \mbox{if}\  i\ \mbox{and}\ j\ \mbox{belong to the same} 
\\
& \mbox{connected component} 
\\[3pt]
0, &  \mbox{otherwise}
\end{array}
\right.
\,    
\label{ao3} 
\end{equation} 
(this should not be mixed with the adjacency matrix).
Then $k_i(t)= \sum_{j=1}^{{\cal N}(t)}S_{ij}(t)$ is the size of a connected component containing $i$-th vertex, that is, the total number of vertices in this component. 

The closed equation can be written for the joint distribution function of  
the 
total number of vertices in the network, ${\cal N}$, the total degree, ${\cal Q} = 2{\cal L}$, and the connected component sizes, $\{ k_i \}$. 
Let us define the $Z$-transform of this joint distribution function as

\begin{equation}
g({\cal N},{\cal Q};x,t) = 
\frac{1}{{\cal N}}\left\langle 
\delta[{\cal N}(t)-{\cal N}]\,\delta[{\cal Q}(t)-{\cal Q}]
\sum_{i=1}^{{\cal N}} x^{k_i(t)}
\right\rangle
\, .   
\label{ao4} 
\end{equation} 
Recall that $\delta(\ )$ is the Kronecker symbol. 
Then, 

\begin{eqnarray}
&& g({\cal N},{\cal Q};x,t+dt) = (1-dt-b\,dt)g({\cal N},{\cal Q};x,t) +
\nonumber
\\[5pt]
&&\frac{dt}{{\cal N}}\sum_{i=1}^{{\cal N}} 
\left\langle 
\delta[{\cal N}(t)+1-{\cal N}]\,\delta[{\cal Q}(t)-{\cal Q}]
x^{k_i(t)}
\right\rangle + 
\nonumber
\\[5pt] 
&&\frac{b\,dt}{{\cal N}} 
\left\langle
\delta[{\cal N}(t)-{\cal N}]\,\delta[{\cal Q}(t)+2-{\cal Q}] 
\times \right.
\nonumber
\\[5pt] 
&& \left.
\sum^{{\cal N}}_{j,l=1}
\frac{[q_j(t)+a][q_l(t)+a]}{[{\cal Q}(t)+a{\cal N}(t)]^2} 
\left\{
\sum_{i=1}^{{\cal N}} x^{k_i}(t) - \right.\right. 
\nonumber
\\[5pt] 
&& 
[1 - S_{il}(t)]
\left[\sum_{m=1}^{{\cal N}} S_{jm}(t)[x^{k_m(t)}-x^{k_j(t)+k_l(t)}] + 
\right.
\nonumber
\\[5pt] 
&& 
\left.\left.\left. 
\sum_{m=1}^{{\cal N}}S_{lm}(t)[x^{k_m(t)}-x^{k_l(t)+k_j(t)}]\right] 
\right\}
\right\rangle 
\, .   
\label{ao5} 
\end{eqnarray}
In the large ${\cal N},{\cal Q}$ limit, one can substitute the factor 
$[1 - S_{il}(t)]$ in Eq. (\ref{ao5}) for $1$, therefore 

\begin{eqnarray}
&& g({\cal N},{\cal Q};x,t+dt) = (1-dt-b\,dt)g({\cal N},{\cal Q};x,t) +
\nonumber
\\[5pt] 
&&
\left(1-\frac{1}{{\cal N}}\right) \frac{1}{{\cal N}-1} 
\left\langle 
\delta[{\cal N}(t)-{\cal N}+1]\,\delta[{\cal Q}(t)-{\cal Q}]
x^{k_i(t)}
\right\rangle + 
\nonumber
\\[5pt] 
&&
\frac{1}{{\cal N}} \langle 
\delta[{\cal N}(t)-{\cal N}+1]\,\delta[{\cal Q}(t)-{\cal Q}]\rangle x + 
\nonumber
\\[5pt] 
&&
b\,dt g({\cal N},{\cal Q}-2;x,t) - 
\nonumber
\\[5pt] 
&& 
\frac{2b\, dt}{{\cal N}({\cal Q}-2+a{\cal N})} 
\left\langle 
\delta[{\cal N}(t)-{\cal N}]\,\delta[{\cal Q}(t)-{\cal Q}+2] \times
\right. 
\nonumber
\\[5pt] 
&& 
\sum_{j=1}^{{\cal N}} [q_j(t)+a] k_j(t) x^{k_j(t)} \rangle + 
\nonumber
\\[5pt] 
&& 
\frac{2b\, dt}{{\cal N}({\cal Q}-2+a{\cal N})^2} 
\left\langle 
\delta[{\cal N}(t)-{\cal N}]\,\delta[{\cal Q}(t)-{\cal Q}+2] \times
\right. 
\nonumber
\\[5pt] 
&& 
\left(\sum^{{\cal N}}_{j=1} [q_j(t)+a] k_j(t) x^{k_j(t)}\right) 
\left(\sum^{{\cal N}}_{l=1} [q_l(t)+a] x^{k_l(t)}\right)
\, .   
\label{ao6} 
\end{eqnarray} 
The sums in Eq. (\ref{ao6}) can be easily calculated: 

\begin{eqnarray}
&& \sum^{{\cal N}(t)}_{i=1} q_i(t) k_i(t) x^{k_i(t)} =
\sum^{{\cal N}(t)}_{i=1} q_i(t) x^{k_i(t)} 
\sum^{{\cal N}(t)}_{j=1} S_{ij}(t) = 
\nonumber
\\[5pt] 
&& 
\sum^{{\cal N}(t)}_{j=1} x^{k_j(t)} 
\sum^{{\cal N}(t)}_{i=1} q_i(t) S_{ij}(t) =
2 \sum^{{\cal N}(t)}_{j=1} [k_i(t)-1] x^{k_j(t)} 
\, .   
\label{ao7} 
\end{eqnarray} 
Here we have used the fact that, in the tree ansatz, loops are absent, so 
$\sum^{{\cal N}(t)}_{i=1} q_i(t) S_{ij}(t) = 2[k_j(t)-1]$. Analogously, 

\begin{equation} 
\sum^{{\cal N}(t)}_{i=1} q_i(t) x^{k_i(t)} = 
2 \sum^{{\cal N}(t)}_{j=1} \left[1-\frac{1}{k_i(t)}\right] x^{k_j(t)} 
\, .   
\label{ao8} 
\end{equation} 
One can see that a general relation

\begin{equation} 
\sum^{{\cal N}(t)}_{i=1} q_i(t) f[k_i(t)] = 
2 \sum^{{\cal N}(t)}_{j=1} \left[1-\frac{1}{k_i(t)}\right] f[k_j(t)] 
\,    
\label{ao9} 
\end{equation} 
holds for arbitrary graphs with tree-like finite components. 
Here $f(\ )$ is an arbitrary function. 
In particular, if $f(k)=1$, then 
$2{\cal Q}(t)=2[{\cal N}(t)-{\cal N}_{com}(t)]$, 
so 
${\cal N}(t) = {\cal Q}(t) + {\cal N}_{com}(t)$, 
where ${\cal N}_{com}(t)$ is the number of components in the network at time $t$. 
Using these relations, we obtain 

\begin{eqnarray}
&& g({\cal N},{\cal Q};x,t+dt) = 
(1-dt-b\,dt)g({\cal N},{\cal Q};x,t) +
\nonumber
\\[5pt] 
&& 
\left(1-\frac{1}{{\cal N}}\right)g({\cal N}-1,{\cal Q};x,t) + 
\frac{1}{{\cal N}}g({\cal N}-1,{\cal Q};1,t)x + 
\nonumber
\\[5pt] 
&& 
b\,dt g({\cal N},{\cal Q}-2;x,t) - 
\nonumber
\\[5pt] 
&& 
\frac{2b\,dt}{{\cal N}({\cal Q}-2+a{\cal N})} 
\left\langle
\delta[{\cal N}(t)-{\cal N}] \delta[{\cal Q}(t)-{\cal Q}+2] \times
\right.
\nonumber
\\[5pt] 
&& 
\left.
\sum^{{\cal N}}_{i=1} [(2+a)k_i(t)-2] x^{k_i(t)}
\right\rangle + 
\nonumber
\\[5pt] 
&& 
\frac{2b\,dt}{{\cal N}({\cal Q}-2+a{\cal N})^2} 
\left\langle
\delta[{\cal N}(t)-{\cal N}] \delta[{\cal Q}(t)-{\cal Q}+2] \times 
\right.
\nonumber
\\[5pt] 
&& 
\left.
\left(\sum^{{\cal N}}_{i=1} [(2+a)k_i(t)-2] x^{k_i(t)}\!\right)\!\!\!
\left(\sum^{{\cal N}}_{j=1} 
\!\left[2+a- \frac{2}{k_i(t)}\right]\!\! x^{k_j(t)}\!\right)
\!\!\right\rangle
.  
\nonumber
\\
[5pt] 
&& \phantom{d} 
\label{ao10} 
\end{eqnarray} 

In particular, when $x=1$, one obtains the joint probability that the network contains ${\cal N}$ vertices and and ${\cal L} = {\cal Q}/2$ edges at time $t$, 

\begin{equation} 
g({\cal N},{\cal Q};1,t) = 
\langle\delta[{\cal N}(t)-{\cal N}]\,
\delta[{\cal Q}(t)-{\cal Q}]  \rangle
\equiv \Pi({\cal N},{\cal Q};t)
\, .   
\label{ao11} 
\end{equation} 
From Eq. (\ref{ao10}), we obtain the equation for $\Pi({\cal N},{\cal Q};t)$: 

\begin{eqnarray}
&& 
\frac{\partial \Pi({\cal N},{\cal Q};t)}{\partial t} = 
\Pi({\cal N}-1,{\cal Q};t) - \Pi({\cal N},{\cal Q};t) +
\nonumber
\\[5pt] 
&& 
\delta[\Pi({\cal N},{\cal Q}-2;t) - \Pi({\cal N},{\cal Q};t)]
\, .   
\label{ao12} 
\end{eqnarray} 
Choosing the initial condition  
$\Pi({\cal N},{\cal Q};t_0) = \delta({\cal N}-{\cal N}_0)
\delta({\cal Q}-{\cal Q}_0)$, we find the solution of Eq. (\ref{ao12}), 

\begin{eqnarray} 
&& \Pi({\cal N},{\cal Q};t) = 
\nonumber
\\[5pt] 
&& 
\frac{t^{{\cal N}-{\cal N}_0}}{({\cal N}-{\cal N}_0)!}\,e^{-t} 
\frac{1+(-1)^{{\cal Q}-{\cal Q}_0}}{2} 
\frac{(b t)^{({\cal Q}-{\cal Q}_0)/2}}{[({\cal Q}-{\cal Q}_0)/2]!}
\,e^{-b t} 
\, .   
\label{ao13} 
\end{eqnarray} 
If we assume that ${\cal Q}_0=2b t_0$ and ${\cal N}_0=t_0$ 
for $t_0 \to \infty$, then Eq. (\ref{ao13}) yields 
$\Pi({\cal N},{\cal Q};t) \to \delta({\cal N}-t)\delta({\cal Q}-2b t)$. 
This shows that the total numbers of vertices and edges, ${\cal N}$ and ${\cal Q}$, are, in fact, rigidly determined in the large network limit.  
Finally, using the decomposition 

\begin{equation} 
\langle x^{k_i} x^{k_j}\rangle \to 
\langle x^{k_i} \rangle\langle x^{k_j} \rangle 
\, ,   
\label{ao14} 
\end{equation}  
which can be justified in the limit of $t \to \infty$, from Eq. (\ref{ao10}), we obtain 

\begin{eqnarray} 
&&  
t\frac{\partial g}{\partial t} + g + 
\frac{2b}{2b+a} [(2+a)x\frac{\partial g}{\partial x} - 2g] -
\nonumber
\\[5pt] 
&& 
\frac{2b}{(2b+a)^2}
[(2+a)x\frac{\partial g}{\partial x} - 2g]
[(2+a)g - 2n] = x
\, ,   
\label{ao15} 
\end{eqnarray} 
where $n(x,t) = \int_0^x dy g(y,t)/y$. 
In the stationary case, Eq. (\ref{ao15}) yields Eq. (\ref{4}) of Sec. \ref{main}. 

If the giant component is absent, 
$g(1,t) = \lim_{x\to 1,{\cal N}\to\infty}g(x,{\cal N};t) = 1$. 
In addition, if all the finite connected components in the network are tree-like, we have 

\begin{eqnarray} 
 n(1,t) & = &
\left\langle\frac{1}{{\cal N}(t)} \sum_{i=1}^{{\cal N}(t)} \frac{1}{k_i(t)}  
\right\rangle = 
\left\langle \frac{{\cal N}_{com}(t)}{{\cal N}(t)} \right\rangle = 
\nonumber
\\[5pt] 
&&
\left\langle \frac{{\cal N}(t)-{\cal Q}(t)/2}{{\cal N}(t)} \right\rangle 
\to 1-b
\, ,   
\label{ao16} 
\end{eqnarray}  
that is, the ``tree condition''. 
The following equation can be written for $n(x,t)$: 

\begin{eqnarray} 
&&  
t\frac{\partial n}{\partial t} + n + 
\frac{2b}{2b+a} [(2+a)g - 2n] -
\nonumber
\\[5pt] 
&& 
\frac{b}{(2b+a)^2}
[(2+a)g - 2n]^2 = x
\, ,   
\label{ao17} 
\end{eqnarray} 

Equation (\ref{ao17}) can be used for the determination of the number of loops in the giant connected component which coincides with the total number ${\cal M}$ of loops in the network since all the finite connected components are tree-like. 
${\cal M}$ indicates the extent of the deviation of the giant component structure from a tree. One sees that ${\cal M}={\cal L}+1-{\cal N}$. 
${\cal L}=bt, {\cal N}=t$, hence the total number of edges in the giant component is equal to 

\begin{eqnarray} 
&&
bt- \sum_k (k-1)N_k(t) = t [b - \sum_k (k-1)n_k(t)] =
\nonumber
\\[5pt] 
&&
t[b-g(1)+n(1)]
\, ,   
\label{ao18} 
\end{eqnarray}  
Subtracting from this expression the total number of vertices in the giant component, that is, $tS=t[1-g(1)]$, we obtain ${\cal M}=t[n(1)-(1-b)]$. It is convenient to introduce 
the relative number of loops for $t \to \infty$: 

\begin{equation} 
\ell \equiv \frac{{\cal M}}{t} = n(1)-(1-b)
\, .   
\label{ao19} 
\end{equation} 
Substituting $n(1)$ from Eq. (\ref{ao19}) into Eq. (\ref{ao17}) taken at the point $x=1$, we obtain the exact equation for $\ell$: 

\begin{equation} 
\ell = \frac{b}{2b+a}[(2+a)S + 2\ell]^2
\, .   
\label{ao20} 
\end{equation} 
This equation also follows from Eq. (\ref{9}). 
Near the phase transition, the giant component is small, so that 

\begin{equation} 
\ell \cong b \frac{(2+a)^2}{2b+a} S^2
\, .   
\label{ao21} 
\end{equation}

\section{Analysis of Eq. (\ref{10})}
\label{aa}

Note, that Eq. (\ref{10}) remains unchanged, if $\zeta \rightarrow 1/\zeta $. 
However, the condition that $\varphi /y\rightarrow
1/\left( 1+\zeta \right) $ at $y\rightarrow \infty $, is not invariant under this change, and for $%
\zeta <1$ we have to choose another solution then for $\zeta >1$. Here we
analyse the solutions of Eq. (\ref{10}) at $y\rightarrow \infty $ and at $%
y\rightarrow 1$. The comparison of both the regions will allow us to
obtain the phase diagram and to find the essential features of the 
probability ${\cal P}(k)$ that a vertex belongs to a connected component of the size $k$.  

If $y\rightarrow \infty $, the term $y^{-1-a}$ on the right-hand side of Eq. (\ref{10})
may be neglected, and we obtain the equation

\begin{equation}
\varphi ^{\prime }\varphi -\varphi =-\frac \zeta {\left( 1+\zeta \right) ^2}y 
\, .
\label{100}
\end{equation}
This equation has two solutions, which are linear in $y$: $\varphi
=y/\left( 1+\zeta \right) $ and $\varphi =\zeta y/\left( 1+\zeta \right) $.
It is the first one, which must be chosen, because it corresponds to $%
g\left( 0\right) =0$. Let us denote the physical solution of Eq. (\ref{10})
as $\varphi _1$, and as $\varphi _2$, --- the other one, which has the 
asymptotic form $\zeta y/\left( 1+\zeta \right) $ at $y\rightarrow \infty $. 
If $%
\zeta >1$, then $\varphi _1\left( y\right) <\varphi _2\left( y\right) $
for $y>1$. This follows from the uniqueness property of a solution of
Eq. (\ref{10}). On the contrary, if $\zeta <1$, the physical solution is a higher
one, $\varphi _1>\varphi _2$.

At $y\rightarrow 1$, after the linearization of the right-hand side of Eq. (\ref{10}%
) with respect to $y-1$ we obtain: 
\begin{equation}
\varphi ^{\prime }\varphi -\varphi =-\left( \frac 14+\beta \right) \left(
y-1\right) \,;\;\beta =\frac \zeta {\left( 1+\zeta \right) ^2}\left(
2+a\right) -\frac 14\,.  \label{110}
\end{equation}
After the substitution $\varphi =\left( 1-y\right) \psi $, Eq. (\ref{110})
takes the form 
 
\[
\left( y-1\right) \psi \frac{d\psi }{dy}=-\left[ \left( \psi -\frac 12%
\right) ^2+\beta \right] \,, 
\]
and can be easily solved: 
\[
\ln \left[ C\left( y-1\right) \right] =-\int \frac{\psi \,d\psi }{\left(
\psi -1/2\right) ^2+\beta }\,. 
\]
Here $C$ is the integration constant. 
The sign of $\beta$ determines a full picture of the set of the solutions of Eq. (\ref
{110}).

At first, let us consider the case $\beta =-\lambda ^2<0$. In this case, 
Eq. (\ref{110}) has 
three families of solutions, real for $y>1$. These families can be written in the implicit form as: 

\begin{eqnarray}
&& C\left( y-1\right)  = 
\nonumber
\\[2pt]
&& \left( \frac \varphi {1-y}-\frac 12+\lambda \right) ^{%
\frac 1{4\lambda }-\frac 12}\left( \frac \varphi {1-y}-\frac 12-\lambda
\right) ^{-\frac 1{4\lambda }-\frac 12}\,,  \label{120a} 
\\[5pt]
&& C\left( y-1\right)  = 
\nonumber
\\[2pt]
&& \left( \frac 12-\lambda -\frac \varphi {1-y}\right) ^{%
\frac 1{4\lambda }-\frac 12}\left( \frac \varphi {1-y}-\frac 12-\lambda
\right) ^{-\frac 1{4\lambda }-\frac 12}\,,  \label{120b} 
\\[5pt]
&& C\left( y-1\right)  = 
\nonumber
\\[2pt]
&& \left( \frac 12-\lambda -\frac \varphi {1-y}\right) ^{%
\frac 1{4\lambda }-\frac 12}\left( \frac 12+\lambda -\frac \varphi {1-y}%
\right) ^{-\frac 1{4\lambda }-\frac 12}\,.  \label{120c}
\end{eqnarray} 
In the family (\ref{120a}), $1/2+\lambda < \varphi(1-y) < +\infty$, 
in (\ref{120b}), $1/2-\lambda < \varphi(1-y) < 1/2+\lambda$, 
in (\ref{120c}), $-\infty < \varphi(1-y) < 1/2-\lambda$.  
Only the families (\ref{120a}) and (\ref{120c}) should be taken into account,
because, for the family (\ref{120b}), we have $g^{\prime \prime }\left( 1\right) <0$.
Here a physical solution is realized in the family (\ref{120a}) if $\zeta <1$, and
in the family (\ref{120c}) when $\zeta >1$. The distinctive feature of the
solutions (\ref{120a}) is that they have nonzero value as $y\rightarrow 1$: $%
\varphi \left( 1\right) =1/C$. This proves that when $\zeta <1$, the giant
component is always present. For the solutions of the family (\ref{120c}) we have $%
\varphi \left( 1\right) =0$, which means the absence of the giant component for $%
\zeta >1$ and $\beta <0$. When $y\rightarrow 1$, we obtain from Eq. (\ref{120c}%
): 

\begin{equation}
\varphi \left( y\right) \approx \left( y-1\right) \left[ \frac 12-\lambda
+C^{\prime }\left( y-1\right) ^{4\lambda /\left( 1-2\lambda \right) }\right]
\,,  
\label{121}
\end{equation} 
where $C^{\prime }$ is some constant of the order of unity if $\lambda \sim 1$.  

If $\beta =\lambda ^2>0$, the solution may be conveniently expressed as: 

\begin{eqnarray}
&& 
C\left( y-1\right) =\left[ \left( \frac \varphi {1-y}-\frac 12\right)
^2+\lambda ^2\right] ^{-1/2} 
\nonumber
\\[5pt]
&& 
\exp \left\{ -\frac 1{2\lambda }\left[ \frac \pi %
2+\arctan \left[ \frac 1\lambda \left( \frac \varphi {1-y}-\frac 12\right)
\right] \right] \right\} 
\,.  
\label{130}
\end{eqnarray}
This form of presentation was chosen 
to ensure
a smooth crossover 
between Eqs. (\ref{120c}) and (\ref{130}) at small $\lambda $. All solutions
of this set take nonzero values as $y\rightarrow 1$:

\begin{equation}
\varphi \left( 1\right) =\frac 1C\exp \left( -\frac \pi {2\lambda }\right)
\,.  \label{131}
\end{equation}
This means, that for $\zeta >1$, the value $\beta =0$ corresponds to the point of the
emergence of the giant connected component in the network. Taking into
account the definition of $\beta $ in Eq. (\ref{110}), we arrive at the
expression (\ref{13}) for the critical line in the $\left( \zeta ,a\right) $
plane. 

Let as consider now the critical region, $\left| \beta \right| \ll 1$. 
If $\beta =0$, the solutions of Eq. (\ref{110}) are given by 
\begin{equation}
\varphi \left( y\right) =\frac{y-1}2\left\{ 1+\frac 1{{\cal W}\left[
-C\left( y-1\right) \right] }\right\} \,,  \label{140}
\end{equation}
where, by definition, the Lambert function ${\cal W}\left( z\right) $ is a proper
solution of the equation ${\cal W}\exp {\cal W}=z$. 
If $-e^{-1}<z<0$, ${\cal W}$ has two real branches: 
the one for which ${\cal W} \to 0$ as $z \to -0$, 
and the other for which ${\cal W} \to -\infty$ in the same limit. 
$z=-e^{-1}$ is the branching point for both these branches. 
When the integration
constant $C$ is positive, one must choose that real negative branch of $%
{\cal W}\left( z\right) $, which tends to $-\infty $ as $y\rightarrow 1+0$.
This ensures a smooth crossover between Eqs. (\ref{120c}) and (\ref{130}) as $%
\lambda \rightarrow 0 $. 

The behavior of the distribution near the critical
line may be treated analytically, if we know the integration constant $%
C=C\left( \zeta \right) $ in Eq. (\ref{140}), that may be obtained by
numerical integration of Eq. (\ref{10}) at the critical line, given by Eq. (%
\ref{13}). Indeed, as $1\gg y-1\gg \exp \left( -\pi /2\lambda \right) $, the
argument of $\arctan $ in Eq. (\ref{130}) becomes negative and large. Then,
using the asymptotic expression: $\arctan z\approx -\pi /2+1/z$, one can see
that the solution (\ref{130}) turns into the solution (\ref{140}) with the same
integration constant $C$. 
Hence, substituting Eq. (\ref{131}) into the
expression for the size of the giant component (\ref{12}), 
accounting for the relation (\ref{13}), and recalling that $\lambda =\sqrt{\beta }$, where $%
\beta $ is defined in Eq. (\ref{110}), one arrives at the expression (\ref{14}%
)  for the giant component size. The resulting factor $D(\zeta)$ in this expression is equal to 
\begin{equation}
D\left( \zeta \right) =\frac{2a\left( \zeta \right) }{C\left( \zeta \right) }%
=\frac{\zeta ^2-6\zeta +1}{2\zeta C\left( \zeta \right) }\,.  \label{150}
\end{equation}

Now let us consider the distribution function for connected components ${\cal P}\left( k\right) $ at the threshold and near it. 
This is the inverse $Z$-transform of $g\left( z\right) $: 

\begin{equation}
{\cal P}\left( k\right) =\oint \frac{dz}{2\pi i}g\left( z\right) z^{-k-1}\,,
\label{160}
\end{equation}
where the integration is performed along the contour around $z=0$, lying
inside the unit circle. After integration by parts, accounting for Eq. (\ref{6}), this expression takes the form 

\begin{equation}
{\cal P}\left( k\right) = \delta(k-1)+\frac{ak}{2\zeta }\oint \frac{dz}{2\pi i}%
h^2\left( z\right) z^{-k-1}\,.  \label{170}
\end{equation}
Introducing the integration variable $y=x^{-1/\left( 2+a\right) }$ 
and $%
\varphi \left( y\right) $, $\varphi /y=h/\left( 1+\zeta \right) $ 
we obtain:
\[
{\cal P}\left( k\right) =\delta(k-1)+a\left( 2+a\right) \frac{\left( 1+\zeta
\right) ^2}{2\zeta }k\int_c\frac{dy}{2\pi i}\varphi ^2\left( y\right)
y^{\left( 2+a\right) k}\,,
\]
where $c$ is some integration contour, lying to the right of $y=1$ point.
At $k\gg 1$, 
the position $y_c$ and character of singularity with the
highest value of $\left| y_c\right| $ 
determine the value of this 
integral. Close to the transition line we have either the singularity at $%
y_c=1$, if $\beta <0$, or at $y_c=1-\varepsilon $, $\varepsilon \ll 1$, if $%
\beta >0$. Hence, when $\lambda =\sqrt{\beta }\gg 1$, and $k$ is large enough,
the vicinity of $y=1$ yields the main contribution to the above
integral. Then one can extend the integration contour to $\pm i\infty $.
Changing the integration variable: $y=1+s$, and assuming that $s$ is small, we
finally obtain the expression for the large $k$ part of the connected component
distribution, 

\begin{equation}
{\cal P}\left( k\right) \approx a\left( 2+a\right) \frac{\left( 1+\zeta \right) ^2}{%
2\zeta }k\int_c\frac{ds}{2\pi i}\varphi ^2\left( 1+s\right) e^{\left(
2+a\right) ks}\,.\,  \label{180}
\end{equation}

At the threshold we have:

\begin{equation}
\varphi ^2\left( 1+s\right) \approx \frac{s^2}4\left( 1+\frac 2{\ln \left(
Cs\right) }\right) \,,  \label{181}
\end{equation}
where we have taken into account, that the appropriate branch of Lambert 
function has the asymptotic form ${\cal W}\left( -z\right) \approx \ln z$ as $%
\left| z\right| \ll 1$. Then, deforming the integration contour to the 
one along the shores of the $s=\left( -\infty ,0\right) $ cut, and calculating
the jump across the cut to the leading order in $1/\ln k$, we obtain:

\begin{equation}
{\cal P}\left( k\right) \approx \frac a{2+a}\frac 1{k^2\ln ^2k}\,.  \label{182}
\end{equation}

When $\beta <0$, that is, in the phase without the giant connected component, and the strong inequality $\lambda \ll 1$ is not valid, we have from Eq. (\ref{121}):

\begin{equation}
\varphi \left( 1+s\right) \approx s^2\left[ \left( \frac 12-\lambda \right)
^2+A_1s^\Delta \right] \,,\;\Delta =\frac{4\lambda }{1-2\lambda }\,,
\label{190}
\end{equation}
with some constant coefficient $A_1\sim 1$. Only the second term in the square brackets is
singular and yields nonzero contribution. Substitution of Eq. (\ref{190})
into Eq. (\ref{180}) yields:

\begin{equation}
{\cal P}\left( k\right) \approx Ak^{-2-\Delta }\,,  
\end{equation}
which is valid if $\ln \left[ \left( 2+a\right) k\right] \gg 1/\lambda $; 
$A$ is a constant. At
smaller $k$ one should use expression for $\varphi \left( 1+s\right) $,
valid at larger $s$. If $\ln s\ll 1/\lambda $, from Eq. (\ref{120c}) the
equation for $\varphi \left( 1+s\right) $ follows: 

\begin{equation}
-Cs=\left[ \frac{2\varphi }s-1\right] ^{-1}\exp \left\{ \left[ \frac{%
2\varphi }s-1\right] ^{-1}\right\} \,,  \label{230}
\end{equation}
whose solution is given by the function $\varphi(y) $, taken precisely at the
threshold, Eq. (\ref{140}). Therefore, as $1\ll \ln \left[ \left( 2+a\right)
k\right] \ll 1/\lambda $, the distribution function assumes the threshold form,
Eq. (\ref{182}).

Above the threshold (i.e., in the phase with the giant connected component), $\beta >0$, the argument of the $\arctan $ function in Eq. (\ref
{130}) is positive and large, and the formula $\arctan z\approx \pi /2-1/z$
may be used. Thus, we obtain the form of the solution:

\begin{equation}
\varphi \left( 1+s\right) \approx \frac s2\left\{ 1+\frac 1{{\cal W}\left[
Cs\exp \left( \frac \pi {2\lambda }\right) \right] }\right\} \,.  \label{240}
\end{equation}
This expression is valid if $\ln \left[ s\exp \left( \frac \pi {2\lambda }%
\right) \right] \ll 1/\lambda $. 
The Lambert function ${\cal W}\left( z\right)$ has a square root type singularity at $z=-e~ {-1}$, which ensures the exponential type
behavior of ${\cal P}\left( k\right) $ at the largest $k$. 
Let us substitute Eq. (\ref
{240}) into Eq. (\ref{180}), retaining only the relevant term.  
Changing
the integration variable, $s=C^{-1}z\exp \left( -\frac \pi {2\lambda }%
+z\right) $, we find the expression for the distribution: 

\begin{eqnarray}
&& 
{\cal P}(k) \approx C^{-3}a\left( 2+a\right) ^2k\exp \left( -\frac{3\pi 
}{2\lambda }\right) 
\nonumber
\\[5pt] 
&&
\int_c\frac{dz}{2\pi i}z\left( z+1\right) \exp \left[
C^{-1}\left( 2+a\right) e^{-\pi /2\lambda }kze^z+3z\right] \,.  \label{250}
\end{eqnarray}
If $\left( 2+a\right) k\gg \exp \left( \frac \pi {2\lambda }\right) $, this
integral can be calculated in the saddle point approximation.
Notice, however,
that the integrand in Eq. (\ref{250}) becomes zero at the saddle point 
$z_c=-1$. To avoid this difficulty, one can perform integration by parts, which
gives: 

\begin{eqnarray}
&&{\cal P}\left( k\right) \approx -C^{-2}a\exp \left( -\frac \pi \lambda
\right)   
\nonumber 
\\[7pt]
&&\int_c\frac{dz}{2\pi i}\left( 2z+1\right) \exp \left[ C^{-1}\left(
2+a\right) e^{-\pi /2\lambda }kze^z+2z\right] \,.  \label{255}
\end{eqnarray}
Then, the saddle point approximation yields: 

\begin{eqnarray} 
&&
{\cal P}\left( k\right) \approx a\left[ 2\pi C^3\left( 2+a\right) ^3k\right]
^{-1/2} 
\nonumber
\\[5pt]
&&
\exp \left[ -\frac{3\pi }{4\lambda }-\frac 32-C^{-1}e^{-\pi /2\lambda
-1}\left( 2+a\right) k\right] \,.  \label{260}
\end{eqnarray}
At smaller values of $k$, but when, nevertheless, it is still possible to use
Eq. (\ref{240}), i.e., when $\left( 2+a\right) k\exp \left( -\frac \pi {%
2\lambda }\right) \ll 1$, but $\left| \ln \left[ \left( 2+a\right) k\exp
\left( -\frac \pi {2\lambda }\right) \right] \right| \ll 1/\lambda $, the
argument of the ${\cal W}$-function in Eq.(\ref{240}) becomes large, and we can
replace the Lambert function with logarithm. 
In the same way as obtaining the threshold distribution (\ref{182}), we 
get the form of the distribution: 

\begin{equation}
{\cal P}\left( k\right) \approx \frac{2a}{2+a}k^{-2}\ln ^{-2}\left[ \left(
2+a\right) k\exp \left( -\frac \pi {2\lambda }\right) \right] \,.
\label{270}
\end{equation}
At even smaller $k$, the expression for $\varphi \left( 1+s\right) $ at
larger $\left| s\right| $ is necessary. It may be found from Eq. (\ref
{130}), if we assume that the argument of the $\arctan$ function is small and
replace $\arctan z$ with $z$. As a result we obtain   

\begin{equation}
\varphi \left( s\right) \approx \frac s2\left[ 1-4\lambda ^2\ln \left(
2\lambda Cse^{\pi /4\lambda }\right) \right] \,,  \label{280}
\end{equation}
which is valid if $\left| \ln \left( \lambda se^{\pi /4\lambda }\right)
\right| \ll 1/\lambda $. In this region the distribution function is of the 
form 

\begin{equation}
{\cal P}\left( k\right) \approx 8\lambda ^2\frac a{2+a}k^{-2}\,.  \label{290}
\end{equation}
Finally, for $k\gg 1$, $\ln k\ll 1/\lambda $, equation (\ref{130}) for $%
\varphi \left( 1+s\right) $ at $s\sim 1/\left[ \left( 2+a\right) k\right] $
assumes the form (\ref{230}), and as a result we obtain the distribution
function in its threshold form, Eq. (\ref{182}).


\section{Another way to get $S(b)$ for the exponential network }
\label{ac}

Here we show how our result for $S(b)$ can be obtained directly for the network growing without preferential attachment of edges, i.e., in the limit $a \to \infty$. 
In this particular case, from Eq. (\ref{3}), we obtain the master equation for the probability ${\cal P}(k)$: 

\begin{eqnarray}
\label{c0}
& & 
t\frac{\partial {\cal P}(k)}{\partial t} + {\cal P}(k) = 
\nonumber
\\[5pt]
& & 
\delta(k-1) +
b k \sum_{j=1}^{k-1}{\cal P}(j){\cal P}(k-j) -
2b k{\cal P}(k)
\, .
\end{eqnarray}  
This is a basic equation for the evolution of the connected components. 
From the long-time limit of Eq. (\ref{c0}), the equation for $g(z)$ follows (see Ref. \cite{chkns01}): 

\begin{equation}
\label{c1}
z g^\prime(z) = \frac{1}{2b}\,\frac{z-g(z)}{1-g(z)} 
\, .
\end{equation}   
The boundary condition for it is $g(0)=0$, so $g^\prime(0)=1/(1+2b)$. 

The threshold solution $g(z,b=1/8)$ approaches $1$ at $z=1$, and $g^\prime(1,b=1/8)=2$. For $b>1/8$, the giant connected component is present, so that, at $z=1$, the corresponding solution of Eq. (\ref{c1}) is less than $1$, and $g^\prime(1)=1/(2b)$. 
When $b<1/8$, i.e., in the phase without the giant connected component, the physical solution $g(z=1)=1$, and $g^\prime(1) = (1-\sqrt{1-8b})/4b$.  
Note that $g(z)$ approaches the point $z=1$ in a non-trivial way. 
Indeed, the values of $g^\prime(z)$ are essentially smaller than 
$g^\prime(1)$ even very close to $z=1$ (see below). 

Near $z=1$, Eq. (\ref{c1}) can be written in the form 

\begin{equation}
\label{c2}
u(\xi) \frac{u(\xi)}{d \xi} = \frac{1}{2b}\,[u(\xi) - \xi]
\, ,
\end{equation} 
where $\xi \equiv 1-z$ and $u \equiv 1-g$. 
Its solution for $b=1/8$, that is, the threshold solution, is 

\begin{equation}
\label{c3}
u(\xi, b=1/8) = 2\xi [1-f(\xi)]
\, ,
\end{equation}  
where $f(\xi)$ is the solution of the transcendental equation 

\begin{equation}
\label{c4}
\ln[\xi f(\xi)] + \frac{1}{f(\xi)} = \ln c
\, .
\end{equation}  
Here, the constant $c=0.295\ldots$ is obtained by the numerical sewing together with the solution $g(z)$ of Eq. (\ref{c1}) passing through $0$ at $z=0$. 

For $b>1/8$, i.e., when the giant connected component is present, the solution of Eq. (\ref{c2}) is given by the following transcendental equation 

\begin{eqnarray}
\label{c5}
- \frac{1}{\sqrt{8b-1}} \arctan \frac{4b[u(\xi)/\xi]-1}{\sqrt{8b-1}} - 
\nonumber
\\[7pt]
\ln\sqrt{\xi^2-u(\xi)\xi+2b u^2(\xi)} = C
\, .
\end{eqnarray}  
One can check that $du(0)/d\xi=1/(2b)$. The constant $C$ is fixed by the value of the solution at $\xi=0$, i.e., $u(0)=S$, where $S$ is the size of the giant connected component: 

\begin{equation}
\label{c6}
C = - \frac{\pi/2}{\sqrt{8b-1}} - \ln \sqrt{2b} - \ln S
\, .
\end{equation} 

When $b$ tends to $1/8$ from above, for $\xi \gg S$, we expand $\arctan$ [accounting that $u(\xi)<2\xi$ in this region] and set $b$ to $1/8$ in the logarithms: 

\begin{eqnarray}
\label{c7}
&& - \frac{\pi/2}{\sqrt{8b-1}} - \ln\frac{1}{2} - \ln S = 
\nonumber
\\[7pt]
&& \frac{\pi/2}{\sqrt{8b-1}} - \frac{1}{1 - u(\xi)/(2\xi)} - \ln[\xi- u(\xi)/2]
\, .
\end{eqnarray}  
We sew together this solution for $b \to 1/8$ and the threshold one (\ref{c3}). One can see that this is possible substituting Eq. (\ref{c3}) into Eq. (\ref{c7}): 

\begin{equation}
\label{c8}
- \frac{\pi}{\sqrt{8b-1}} + \frac{1}{f(\xi)} - 
\ln \frac{S}{2} = - \ln[\xi f(\xi)]
\, .
\end{equation} 
Accounting for Eq. (\ref{c4}), we finally obtain  

\begin{equation}
\label{c9}
S = 2c \exp\left[-\frac{\pi}{2\sqrt{2}}\,\frac{1}{\sqrt{b-1/8}}\right]
\, ,
\end{equation} 
where the coefficient $2c=0.590\ldots$. 

One should note that the accurate sewing procedure has been necessary only for the determination of the coefficient of the exponent in Eq. (\ref{c9}). Indeed, the index of the exponent can be easily obtained without consideration of the last two terms on the right-hand side of Eq. (\ref{c7}) for $g(x)$.



\vspace{33mm}$\phantom{xxx}$

\vspace{33mm}$\phantom{xxx}$

\vspace{33mm}$\phantom{xxx}$

\end{multicols}

\end{document}